# Alchemi: A .NET-based Grid Computing Framework and its Integration into Global Grids

Akshay Luther, Rajkumar Buyya, Rajiv Ranjan, and Srikumar Venugopal

**Gri**d Computing and **D**istributed **S**ystems (GRIDS) Laboratory
Department of Computer Science and Software Engineering
The University of Melbourne, Australia
Email:{akshayl, raj, rranjan, srikumar}@cs.mu.oz.au

**Abstract**: Computational grids that couple geographically distributed resources are becoming the de-facto computing platform for solving large-scale problems in science, engineering, and commerce. Software to enable grid computing has been primarily written for Unix-class operating systems, thus severely limiting the ability to effectively utilize the computing resources of the vast majority of desktop computers i.e. those running variants of the Microsoft Windows operating system. Addressing Windows-based grid computing is particularly important from the software industry's viewpoint where interest in grids is emerging rapidly. Microsoft's .NET Framework has become near-ubiquitous for implementing commercial distributed systems for Windows-based platforms, positioning it as the ideal platform for grid computing in this context. In this paper we present Alchemi, a .NET-based grid computing framework that provides the runtime machinery and programming environment required to construct desktop grids and develop grid applications. It allows flexible application composition by supporting an object-oriented grid application programming model in addition to a grid job model. Cross-platform support is provided via a web services interface and a flexible execution model supports dedicated and non-dedicated (voluntary) execution by grid nodes.

## 1 Introduction

The idea of metacomputing [2] is very promising as it enables the use of a network of many independent computers as if they were one large parallel machine, or virtual supercomputer for solving large-scale problems in science, engineering, and commerce. With the exponential growth of global computer ownership, local networks and Internet connectivity, this concept has been taken to a new level - grid computing [1][8].

There is rapidly emerging interest in grid computing from commercial enterprises. A Microsoft Windows-based grid computing infrastructure will play a critical role in the industry-wide adoption of grids [9][14] [17][22] due to the large-scale deployment of Windows within enterprises. This enables the harnessing of the unused computational power of desktop PCs and workstations to create a virtual supercomputing resource at a fraction of the cost of traditional supercomputers.  However, there is a distinct lack of service-oriented architecture-based grid computing software in this space. To overcome this limitation, we have developed a Windows-based grid computing framework called Alchemi implemented on the Microsoft .NET Platform.

While the notion of grid computing is simple enough, the practical realization of grids poses a number of challenges. Key issues that need to be dealt with are security, heterogeneity, reliability, application composition, scheduling, and resource management [13]. The Microsoft .NET Framework [3] provides a powerful toolset that can be leveraged for all of these, in particular support for remote execution (via .NET Remoting [4] and web services [23]), multithreading, security, asynchronous programming, disconnected data access, managed execution and cross-language development, making it an ideal platform for grid computing middleware.

The Alchemi grid computing framework was conceived with the aim of making grid construction and development of grid software as easy as possible without sacrificing flexibility, scalability, reliability and extensibility. The key features supported by Alchemi are:

- Internet-based clustering [20][21] of desktop computers without a shared file system;
- federation of clusters to create hierarchical, cooperative grids;
- dedicated or non-dedicated (voluntary) execution by clusters and individual nodes;
- object-oriented grid thread programming model (fine-grained abstraction); and
- web services interface supporting a grid job model (coarse-grained abstraction) for cross-platform interoperability e.g. for creating a global and cross-platform grid environment via a custom resource broker component.

A scenario for the creation of a layered architecture-based grid computing environment using Alchemi and other grid technologies such as Globus Toolkit [7] is shown in Figure 1. The Gridbus Grid Service Broker (GSB), originally designed to operate with resources grid-enabled using Globus, has been extended to operate with resources grid-enabled using Alchemi via Alchemi's cross-platform web services interface. In such an environment, grid applications can be created using either the grid thread model supported by Alchemi or parameter-sweep programming model supported by the Gridbus broker. Applications designed using Alchemi's object-oriented grid thread model (written in a .NET-supported language) run on Alchemi nodes, whereas applications formulated as parameter-sweep applications can be deployed on either Alchemi or Globus nodes as long as there exist executables for all target platforms.

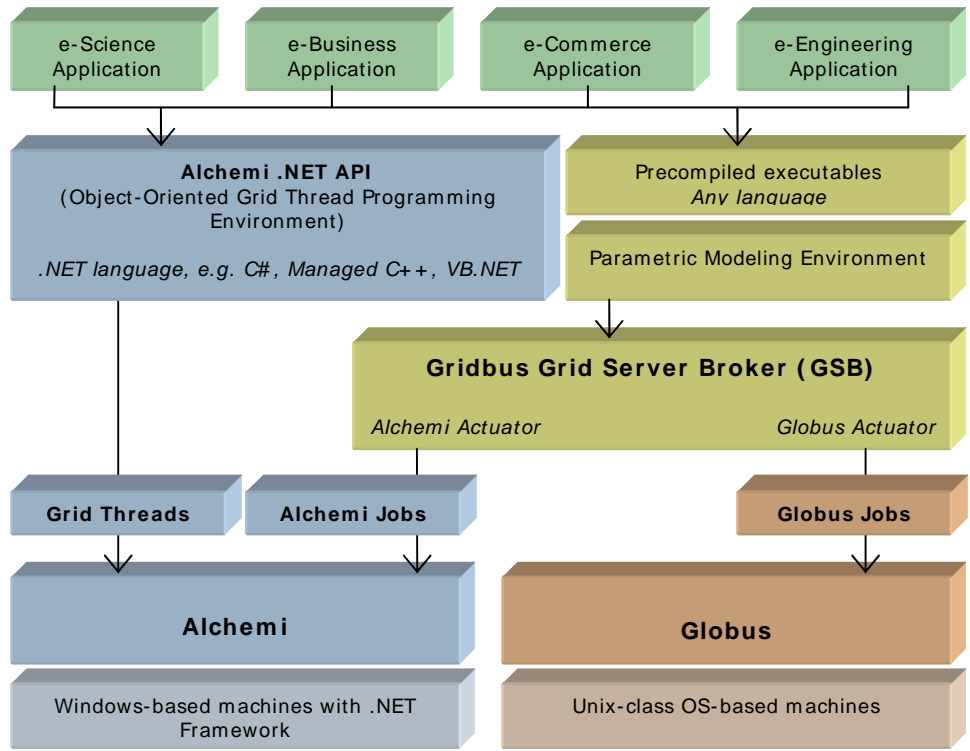

**Figure 1. A layered architecture for integration of distributed Windows and Unix-class resources.**

The rest of the paper is organized as follows. Section 2 presents the Alchemi architecture and discusses configurations for creating different grid environments. Section 3 discusses the system implementation and presents the lifecycle of an Alchemi-enabled grid application demonstrating its execution model. Section 4 presents the object-oriented grid thread programming model supported by the Alchemi API. Section 5 presents the results of an evaluation of Alchemi as a platform for execution of applications written using the Alchemi API. It also evaluates the use of Alchemi nodes as part of a global grid alongside Unix-class grid nodes running Globus software. Section 5 presents related works along with their comparison to Alchemi. Finally, we conclude the paper with work planned for the future.

# 2 Architecture

Alchemi follows the master-worker parallel programming paradigm [30] in which a central component dispatches independent units of parallel execution to workers and manages them. This smallest unit of parallel execution is a grid thread, which is conceptually and programmatically similar to a thread object (in the object-oriented sense) that wraps a "normal" multitasking operating system thread. A grid application is defined simply as an application that is to be executed on a grid and that consists of a number of grid threads. Grid applications and grid threads are exposed to the grid application developer via the object-oriented Alchemi .NET API (see section 4, *Alchemi API: Grid Thread Programming Model*).

## 2.1 Components

Alchemi offers four distributed components as illustrated in Figure 2, designed to operate under three usage patterns. They are discussed here with respect to their basic operation.

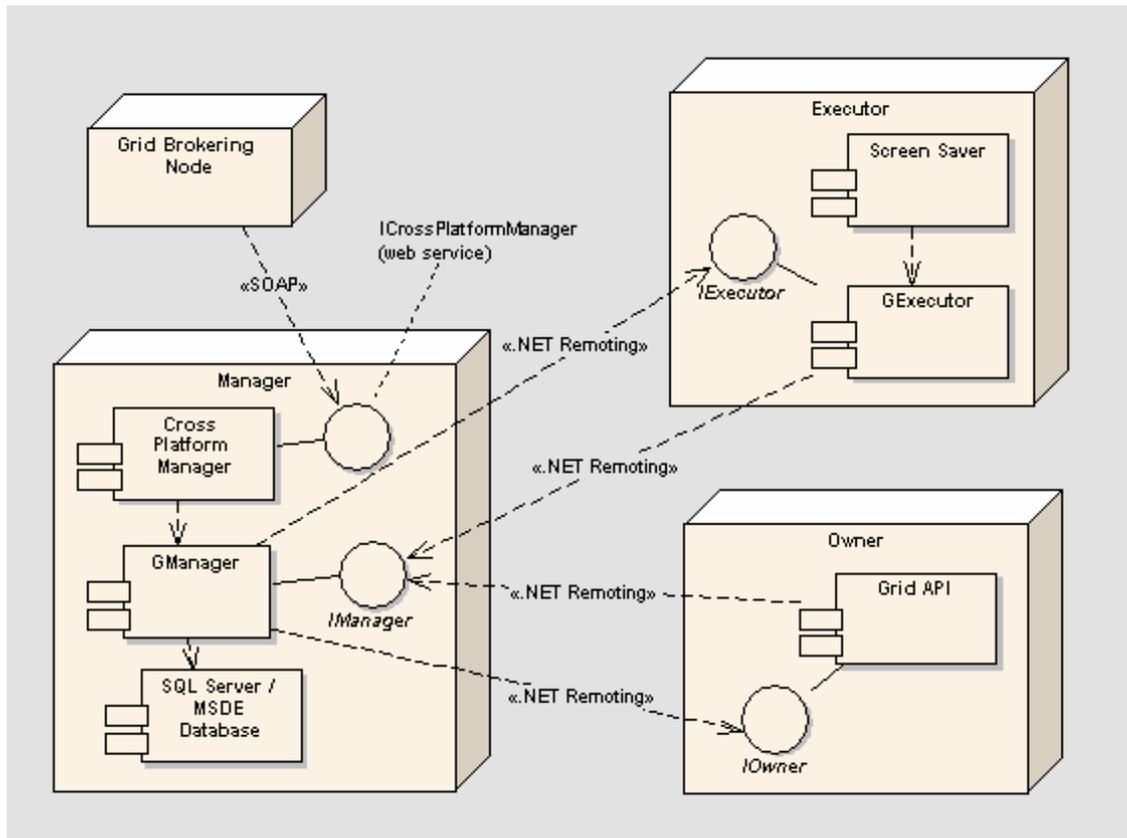

**Figure 2. Alchemi architecture and interaction between its components.**

**Manager**
The Manager manages the execution of grid applications and provides services associated with managing thread execution. The Executors register themselves with the Manager which in turn keeps track of their availability. Threads received from the Owner are placed in a pool and scheduled to be executed on the various available Executors. A priority for each thread can be explicitly specified when it is created within the Owner, but is assigned the highest priority by default if none is specified. Threads are scheduled on a Priority and First Come First Served (FCFS) basis, in that order. The Executors return completed threads to the Manager which are subsequently passed on or collected by the respective Owner.

**Executor**
The Executor accepts threads from the Manager and executes them. An Executor can be configured to be dedicated, meaning the resource is centrally managed by the Manager, or non-dedicated, meaning that the resource is managed on a volunteer basis via a screen saver or by the user. For non-dedicated execution, there is one-way communication between the Executor and the Manager. In this case, the resource that the Executor resides on is managed on a volunteer basis since it requests threads to execute from the Manager. Where two-way communication is possible and dedicated execution is desired the Executor exposes an interface (**IExecutor**) so that the Manager may communicate with it directly. In this case, the Manager explicitly instructs the Executor to execute threads, resulting in centralized management of the resource where the Executor resides. Thus, Alchemi's execution model provides the dual benefit of:

- flexible resource management i.e. centralized management with dedicated execution vs. decentralized management with non-dedicated execution; and
- flexible deployment under network constraints i.e. the component can be deployment as non-dedicated where two-way communication is not desired or not possible (e.g. when it is behind a firewall or NAT/proxy server).

Thus, dedicated execution is more suitable where the Manager and Executor are on the same Local Area Network while non-dedicated execution is more appropriate when the Manager and Executor are to be connected over the Internet.

**Owner**
Grid applications created using the Alchemi API are executed on the Owner component. The Owner provides an interface with respect to grid applications between the application developer and the grid. Hence it "owns" the application and provides services associated with the ownership of an application and its constituent threads. The Owner submits threads to the Manager and collects completed threads on behalf of the application developer via the Alchemi API.

**Cross-Platform Manager**
The Cross-Platform Manager, an optional sub-component of the Manager, is a generic web services interface that exposes a portion of the functionality of the Manager in order to enable Alchemi to manage the execution of platform independent grid jobs (as opposed to grid applications utilizing the Alchemi grid thread model). Jobs submitted to the Cross-Platform Manager are translated into a form that is accepted by the Manager (i.e. grid threads), which are then scheduled and executed as normal in the fashion described above. Thus, in addition to supporting the grid-enabling of existing applications, the Cross-Platform Manager enables other grid middleware to interoperate with and leverage Alchemi on any platform that supports web services (e.g. Gridbus Grid Service Broker).

## 2.2 System Configurations

The components discussed above allow Alchemi to be utilized to create different grid configurations: desktop cluster grid, multi-cluster grid, and cross-platform grid (global grid).

**Cluster (Desktop Grid)**
The basic deployment scenario – a cluster (shown in Figure 3) - consists of a single Manager and multiple Executors that are configured to connect to the Manager. One or more Owners can execute their applications on the cluster by connecting to the Manager. Such an environment is appropriate for deployment on Local Area Networks as well as the Internet. The operation of the Manager, Executor and Owner components in a cluster is as described above.

**Multi-Cluster**
A multi-cluster environment is created by connecting Managers in a hierarchical fashion (Figure 34a). As in a single-cluster environment, any number of Executors and Owners can connect to a Manager at any level in the hierarchy. An Executor and Owner in a multi-cluster environment connect to a Manager in the same fashion as in a cluster and correspondingly their operation is no different from that in a cluster.

The key to accomplishing multi-clustering in Alchemi's architecture is the fact that a Manager behaves like an Executor towards another Manager since the Manager implements the interface of the Executor. A Manager at each level except for the topmost level in the hierarchy is configured to connect to a higher-level Manager as an "intermediate" Manager and is treated by the higher level-Manager as an Executor. Such an environment is more appropriate for deployment on the Internet.

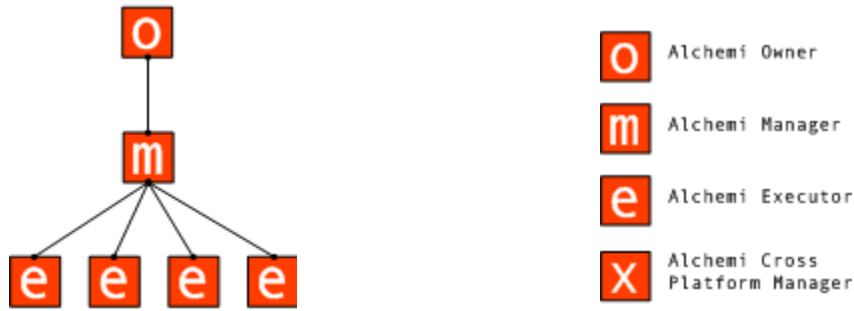

[a] Cluster (desktop grid) deployment  [b] Legend

**Figure 3. Cluster (desktop grid) deployment.**

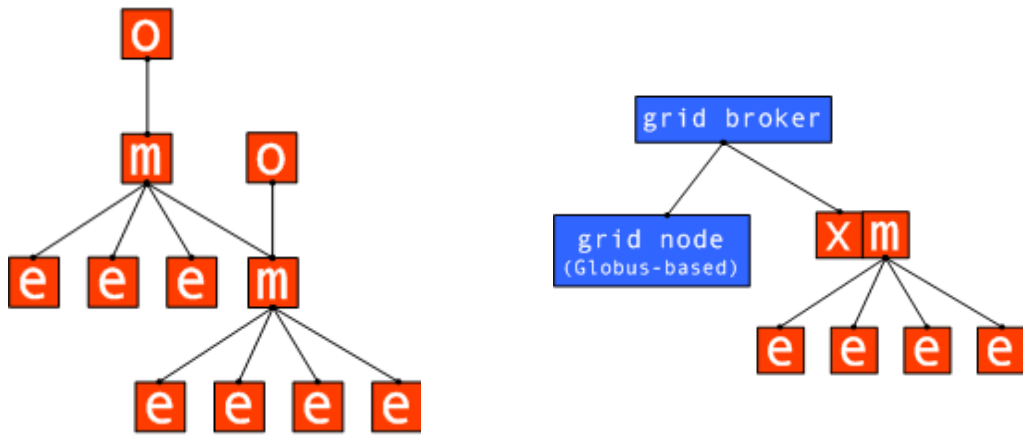

[a] Multi-cluster deployment  [b] Cross-platform (global grid) deployment

**Figure 4. Alchemi deployment in [a] multi-cluster and [b] global grid environments.**

The operation of an intermediate Manager in a multi-cluster environment therefore, must be discussed with respect to the behavior of an Executor and is as follows.

Once Owners have submitted grid applications to their respective Managers, each Manager has "local" grid threads waiting to be executed. As discussed, threads are assigned the highest priority by default (unless the priority is explicitly specified during creation) and threads are scheduled and executed as normal by the Manager's local Executors. Note that an 'Executor' in this context could actually be an intermediate Manager, since it is treated as an Executor by the higher-level Manager. In this case after receiving a thread from the higher-level Manager, it is scheduled locally by the intermediate Manager with a priority reduced by one unit and is executed as normal by the Manager's local 'Executors' (again, any of which could be intermediate Managers).

In addition, at some point the situation may arise when a Manager wishes to allocate a thread to one of its local Executors (one or more of which could an intermediate Manager), but there are no local threads waiting to be executed. In this case, if the Manager is an intermediate Manager, it requests a thread from its higher-level Manager, reduced the priority by one unit and schedules it locally.

In both of these cases, the effect of the reduction in priority of a thread as it moves down the hierarchy of Managers is that the "closer" a thread is submitted to an Executor, the higher is the priority that it executes with. This allows a portion of an Alchemi grid that is within one administrative domain (i.e. a cluster or multi-cluster under a specific "administrative domain Manager") to be shared with other organizations to create a collaborative grid environment without impacting on its utility to local users.

As with an Executor, an intermediate Manager must be configured for either dedicated or non-dedicated execution. Not only does its operation in this respect mirror that of an Executor, the same benefits of flexible resource management and deployment under network constraints apply.

**Cross-Platform Grid**

The Cross-Platform Manager can be used to construct a grid conforming to the classical global grid model (Figure 34b). A grid middleware component such as a broker can use the Cross-Platform Manager web service [25][26][27][28] to execute cross-platform applications (jobs within tasks) on an Alchemi node (cluster or multi-cluster) as well as resources grid-enabled using other technologies such as Globus.

## 3    Design and Implementation

Figure 5 provide an overview of the implementation by way of a class diagram (showing only the main classes without attributes or operations).

### 3.1    Overview

The .NET Framework offers two mechanisms for execution across application domains – Remoting and web services (application domains are the unit of isolation for a .NET application and can reside on different network hosts).

.NET Remoting allows a .NET object to be "remoted" and expose its functionality across application domains. Remoting is used for communication between the four Alchemi distributed grid components as it allows low-level interaction transparently between .NET objects with low overhead (remote objects are configured to use binary encoding for messaging).

Web services were considered briefly for this purpose, but were decided against due to the relatively higher overheads involved with XML-encoded messages, the inherent inflexibility of the HTTP protocol for the requirements at hand and the fact that each component would be required to be configured with a web services container (web server). However, web services are used for the Cross-Platform Manager's public interface since cross-platform interoperability was the primary requirement in this regard.

The objects remoted using .NET Remoting within the four distributed components of Alchemi, the Manager, Executor, Owner and Cross-Platform Manager are instances of **GManager**, **GExecutor**, **GApplication** and **CrossPlatformManager** respectively.

**GManager**, **GExecutor**, **GApplication** derive from the **GNode** class which implements generic functionality for remoting the object itself and connecting to a remote Manager via the **IManager** interface.

The Manager executable initializes an instance of the **GManager** class, which is always remoted and exposes a public interface **IManager**. A key point to note is the fact the **IManager** interface derives from **IExecutor**. This allows a Manager to connect to another Manager and appear to be an Executor. This is the means by which the architecture supports the building of hierarchical grids.

The Executor executable creates an instance of the **GExecutor** class. For non-dedicated execution, there is one-way communication between the Executor and the Manager. Where two-way communication is possible and dedicated execution is desired, **GExecutor** is remoted and exposes the **IExecutor** interface so that the Manager may communicate with it directly. The Executor installation provides an option to install a screen saver, which initiates non-dedicated execution when activated by the operating system.

The Owner component - inside which instances of the **GApplication** object are created via Alchemi API - communicates with the Manager in a similar fashion to **GExecutor**. While two-way communication is currently not used in the implementation, the architecture caters for this by way of the **IOwner** interface.

The Cross-Platform Manager web service is a thin wrapper around **GManager** and uses applications and threads internally to represent tasks and jobs (the **GJob** class derives from **GThread**) via the public **ICrossPlatformManager** interface.

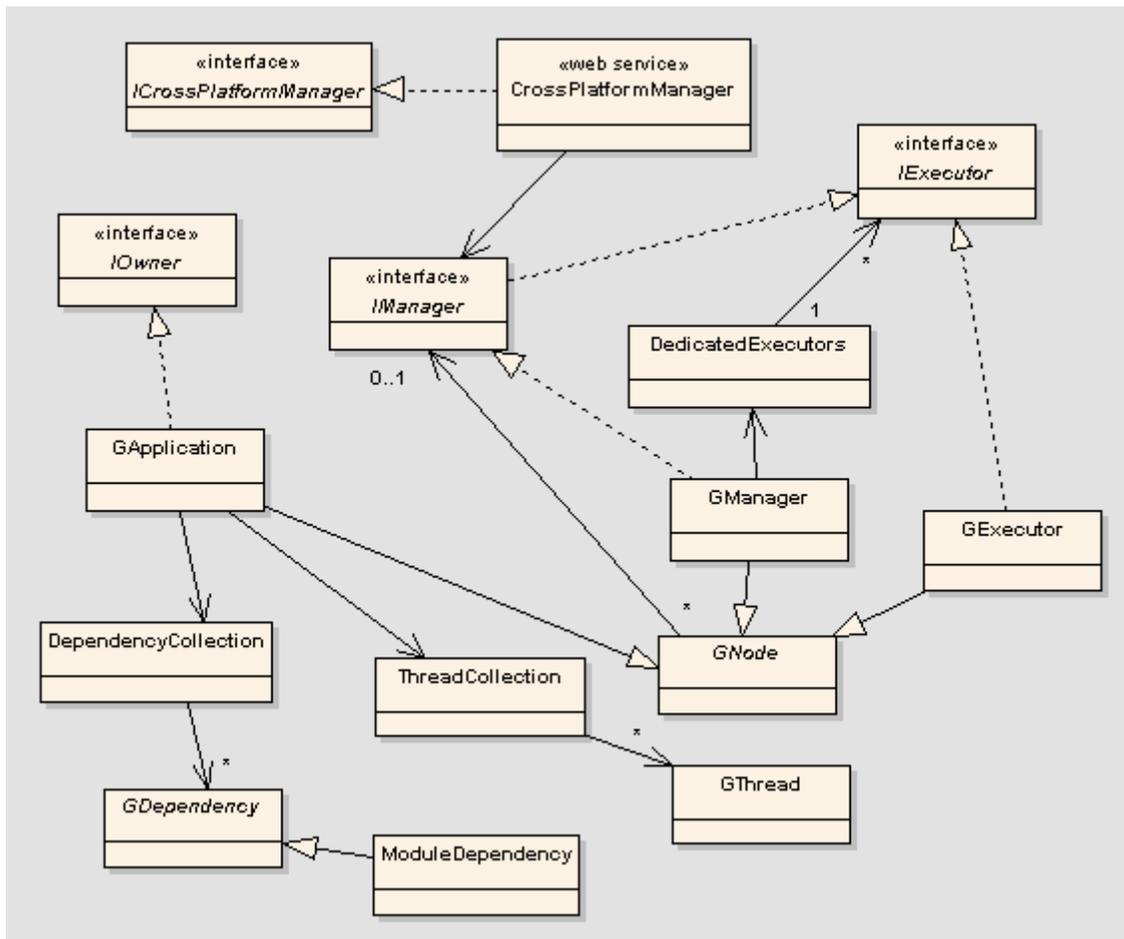

**Figure 5. Main classes and their relationships.**

## 3.2  Grid Application Lifecycle

To develop and execute a grid application the developer creates a custom grid thread class that derives from the abstract **GThread** class. An instance of the **GApplication** object is created and any dependencies required by the application are added to its **DependencyCollection**. Instances of the **GThread**-derived class are then added to the **GApplication**'s **ThreadCollection**.

The lifecycle of a grid application is shown in Figure 6 and Figure 7, showing simplified interactions between the Owner and Executor nodes respectively and the Manager.

The **GApplication** serializes and sends relevant data to the Manager, where it is persisted to disk and threads scheduled. Application and thread state is maintained in a SQL Server / MSDE database.

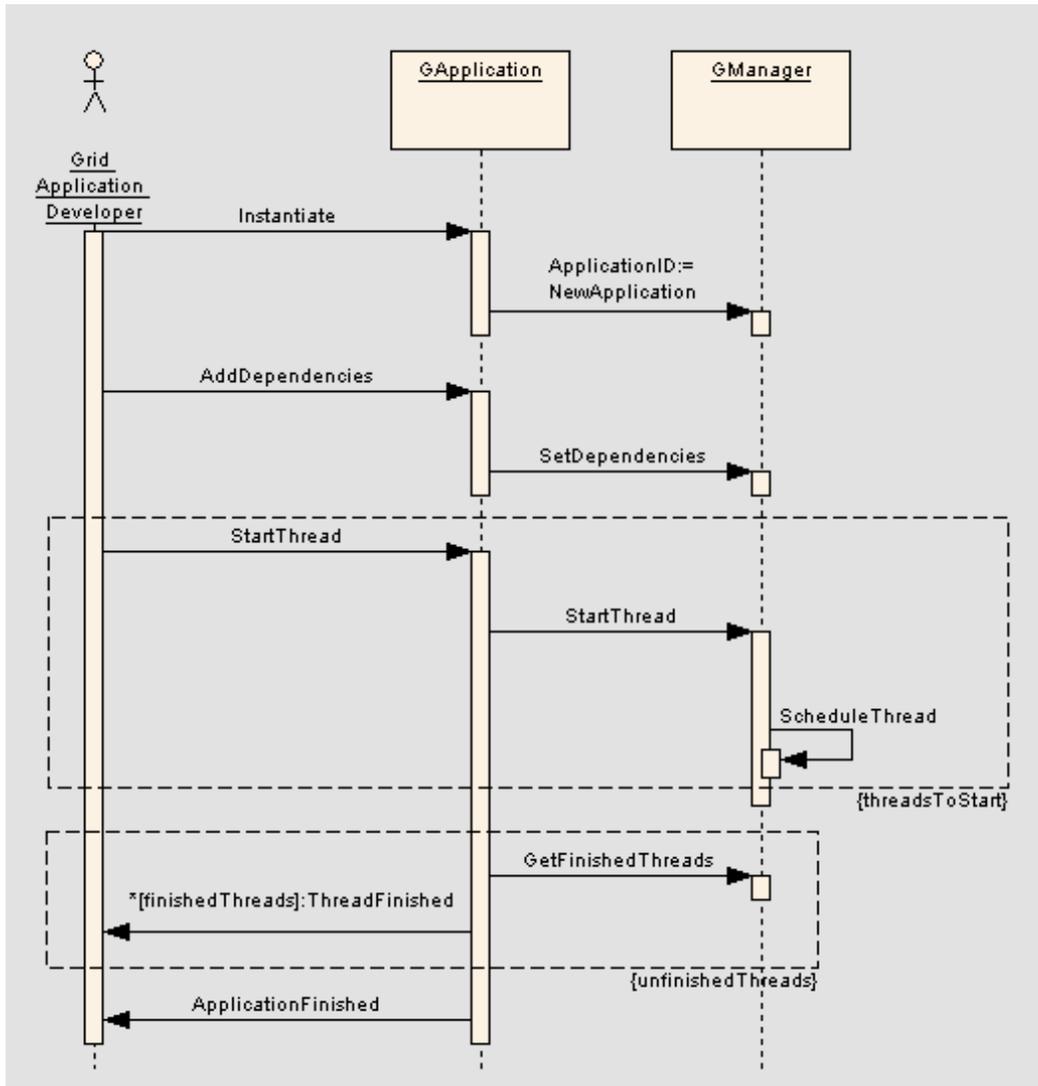

**Figure 6. Simplified interaction between Owner and Manager nodes.**

Non-dedicated executors poll for threads to execute until one is available. Dedicated executors are directly provided a thread to execute by the Manager.

Threads are executed in .NET application domains, with one application domain for each grid application. If an application domain does not exist that corresponds to the grid application that the thread belongs to, one is created by requesting, deserilizing and dynamically loading the application's dependencies. The thread object itself is then desterilized, started within the application domain and returned to the Manager on completion.

After sending threads to the Manager for execution, the **GApplication** polls the Manager for finished threads. A user-defined **GThreadFinish** delegate is called to signify each thread's completion and once all threads have finished a user-defined **GApplicationFinish** delegate is called.

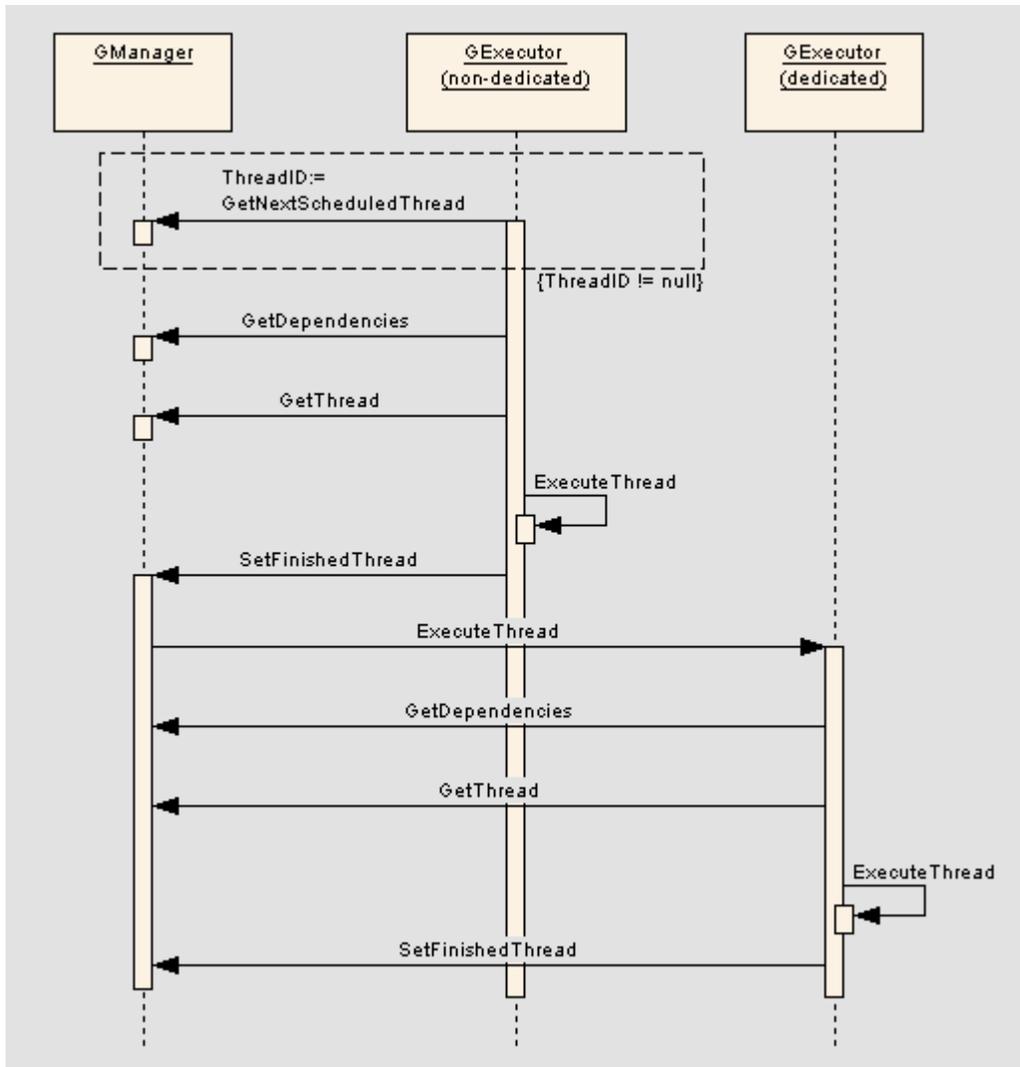

**Figure 7. Simplified interaction between Executor and Manager nodes.**

## 4  Alchemi API: Grid Thread Programming Model

### 4.1  Introduction

**Grid Thread Model**
Alchemi simplifies the development of grid applications by providing a programming model that is object-oriented and that imitates traditional multi-threaded programming. The atomic unit of parallel execution is a grid thread with many grid threads comprising a grid application (hereafter, 'applications' and 'threads' can be taken to mean grid applications and grid threads respectively, unless stated otherwise).

Developers deal only with application and thread objects and any other custom objects, allowing him/her to concentrate on the grid application itself without worrying about the "plumbing" details. Furthermore, this kind of abstraction allows the use of programming language constructs such as events between local and remote code. All of this is offered via the Alchemi .NET API.

The additional benefit of this approach is that it does not limit the developer to applications that are completely or "embarrassingly" parallel. Indeed, it allows development of grid applications where inter-thread communication is required. While Alchemi currently only supports completely parallel threads, support for inter-thread communication is planned for the future. Finally it should be noted that grid applications utilizing the Alchemi .NET API can be written in any .NET-supported language e.g. C#, VB.NET, Managed C++, J#, JScipt.NET.

**Grid Job Model**

Traditional grid implementations have offered a high-level abstraction of the "virtual machine", where the smallest unit of parallel execution is a process (typically referred to as a job, with many jobs constituting a task). Although writing software for the "grid job" model involves dealing with processes, an approach that can be complicated and inflexible, Alchemi's architecture supports it via web services interface for the following reasons:

- grid-enabling existing applications; and
- cross-platform interoperability with grid middleware that can leverage Alchemi

Grid tasks and grid jobs are represented internally as grid applications and grid threads respectively.

## 4.2   Overview of Grid Application Development with Alchemi

The two central classes in the Alchemi .NET API are **GThread** and **GApplication**, representing a grid thread and grid application respectively. There are essentially two parts to an Alchemi grid application. Each is centered on one of these classes:

- code to be executed remotely (a grid thread and its dependencies) and
- code to be executed locally (the grid application itself).

A custom grid thread is implemented by writing a class that derives from **GThread**, overriding the **void Start()** method, and marking the class with the **Serializable** attribute. Code to be executed remotely is defined in the implementation of the overridden **Start** method. This **GThread**-derived class and any dependencies (that are not part of the .NET Framework) must be compiled as one or more .NET Assemblies. Figure 8 shows a very simple grid thread that multiplies two integers.

```csharp
using System;
using Alchemi.Core;
namespace Alchemi.Examples.Tutorial
{
  [Serializable]
  public class MultiplierThread : GThread
  {
    private int _A;
    private int _B;
    private int _Result;
    public int Result
    {
      get { return _Result; }
    }
    public MultiplierThread(int a, int b)
    {
      _A = a;
      _B = b;
    }
    public override void Start()
    {
      _Result = _A * _B;
    }
  }
}
```

**Figure 8. Listing of a simple grid thread (C#).**

The grid application itself can be any type of .NET application. It creates instances of the custom grid thread, executes them on the grid and uses each thread's results. Figure 9 shows a very simple grid application that uses the custom **MutliplierThread** grid thread discussed above.

```csharp
using System;
using Alchemi.Core;

namespace Alchemi.Examples.Tutorial
{
  class GridApplication
  {
    [STAThread]
    static void Main(string[] args)
    {
      Console.WriteLine("[enter] to start grid application ...");
      Console.ReadLine();
      // create grid application
      GApplication ga = new GApplication("localhost", 9000);
      // add GridThread module as a dependency
      ga.Manifest.Add(
        new ModuleDependency(typeof(MultiplierThread).Module)
      );
      // create and add 10 threads to the application
      for (int i=0; i<10; i++)
      {
        // create thread
        MultiplierThread thread = new MultiplierThread(i, i+1);

        // set the thread finish callback method
        thread.FinishCallback = new GThreadFinish(ThreadFinished);

        // add thread to application
        ga.Threads.Add(thread);
      }
      // set the application finish callback method
      ga.FinishCallback = new GApplicationFinish(ApplicationFinished);
      // start application
      ga.Start();
      Console.ReadLine();
    }
    static void ThreadFinished(GThread th)
    {
      // cast GThread back to MultiplierThread
      MultiplierThread thread = (MultiplierThread) th;
      Console.WriteLine(
        "thread # {0} finished with result '{1}'", thread.Id, thread.Result);
    }
    static void ApplicationFinished()
    {
      Console.WriteLine("\napplication finished\n");
      Console.WriteLine("[enter] to continue ...");
    }
  }
}
```

**Figure 9. Listing of a simple grid application (C#).**

# 5   Performance Evaluation

## 5.1   Standalone Alchemi Grid

**Testbed**

The testbed is an Alchemi cluster consisting of six Executors (Pentium III 1.7 GHz desktop machines with 512 MB physical memory running Windows 2000 Professional). One of these machines is additionally designated as a Manager.

**Test Application & Methodology**

The test application is the computation of the value of Pi to n decimal digits. The algorithm used allows the computation of the p'th digit without knowing the previous digits [28]. The application utilizes the Alchemi grid thread model (described in section 4.2, *Alchemi API: Grid Thread Programming Model*). The test was performed for a range of workloads (calculating 1000, 1200, 1400, 1600, 1800, 2000 and 2200 digits of Pi), each with one to six Executors enabled. The workload was sliced into a number of threads, each to calculate 50 digits of Pi, with the number of threads varying proportionally with the total number of digits to be calculated. Execution time was measured as the elapsed clock time for the test program to complete on the Owner node.

**Results**

Figure 10 shows a plot between thread size (the number of decimal places to which Pi is calculated to) and total time (in seconds taken by the all threads to complete execution) with varying numbers of Executors enabled.

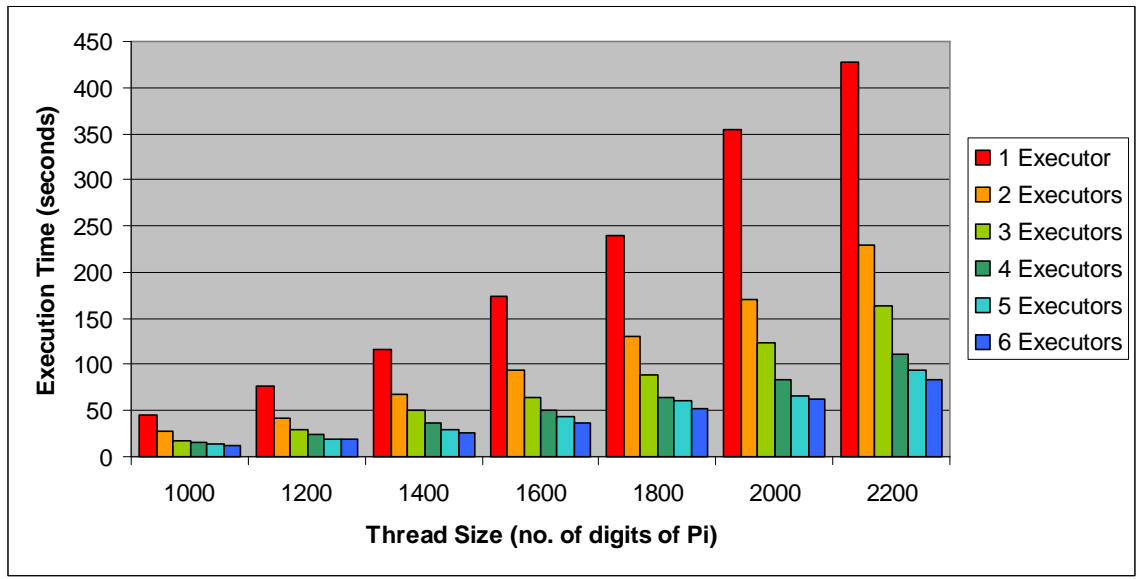

**Figure 10. A plot of thread size vs. execution time on a standalone Alchemi cluster with varying numbers of Executors enabled.**

At a low workload (1000 digits), there is little difference between the total execution time with different quantity of Executors. This is explained by the fact that the total overhead (network latency and miscellaneous overheads involved in managing a distributed execution environment) is in a relatively high proportion to the actual total computation time. However, as the workload is increased, there is near-proportional difference when higher numbers of executors are used. For example, for 2200 digits, the execution time with six executors (84 seconds) is nearly 1/5$^{th}$ of that with one executor (428 seconds). This is explained by the fact that for higher workloads, the total overhead is in a relatively lower proportion to the actual total computation time.

## 5.2 Cross-Platform Global Grid

**Testbed**

A global grid was used for evaluating Alchemi in a cross-platform environment, with the Gridbus Grid Service Broker managing five grid resources (Table 1). One of the resources was the cluster of Alchemi nodes described in the previous section while the other resources ran Globus 2.4 [7]. The Gridbus resource brokering mechanism used in this test obtains the users' application requirements and evaluates the suitability of various resources for the purpose. It then schedules the jobs to various resources in order to satisfy those requirements.

| Resource | Location | Configuration | Grid Middleware | Jobs Completed |
|---|---|---|---|---|
| maggie.cs.mu.oz.au [Windows cluster] | University of Melbourne | 6 * Intel Pentium IV 1.7 GHz | Alchemi | 21 |
| quidam.ucsd.edu [Linux cluster] | University of California, San Diego | 1 * AMD Athlon XP 2100+ | Globus | 16 |
| belle.anu.edu.au [Linux cluster] | Australian National University | 4 * Intel Xeon 2 | Globus | 22 |
| koume.hpcc.jp [Linux cluster] | AIST, Japan | 4 * Intel Xeon 2 | Globus | 18 |
| brecca-2.vpac.org [Linux cluster] | VPAC Melbourne | 4 * Intel Xeon 2 | Globus | 23 |

**Table 1. Grid resources.**

**Test Application & Methodology**

For the purpose of evaluation, we used an application that calculates mathematical functions based on the values of two input parameters. The first parameter *X*, is an input to a mathematical function and the second parameter *Y*, indicates the expected calculation complexity in minutes plus a random deviation value between 0 to 120 seconds—this creates an illusion of small variation in execution time of different parametric jobs similar to a real application. A plan file modeling this application as a parameter sweep application using the Nimrod-G parameter specification language [12] is shown in Figure 11. The first part defines parameters and the second part defines the task that is to be performed for each job. As the parameter *X* varies from values 1 to 100 in step of 1, this plan file would create 100 jobs with input values from 1 to 100.

```
#Parameter definition
parameter X integer range from 1 to 100 step 1;
parameter Y integer default 1;
#Task definition
task main
        #Copy necessary executables depending on node type
        copy calc.$OS node:calc
        #Execute program with parameter values on remote node
        node:execute ./calc $X $Y
        #Copy results file to use home node with jobname as extension
        copy node:output ./output.$jobname
endtask
```

**Figure 11. Parametric job specification.**

**Results**

The results of the experiment in Figure 12 show the number of jobs completed on different Grid resources at different times. The parameter calc.$OS directs the broker to select appropriate executable based a target Grid resource architecture. For example, if the target resource is Windows/Intel, it selects calc.exe and copies to the grid node before its execution. It demonstrates the feasible to utilizing Windows-based Alchemi resources along with other Unix-class resources running Globus.

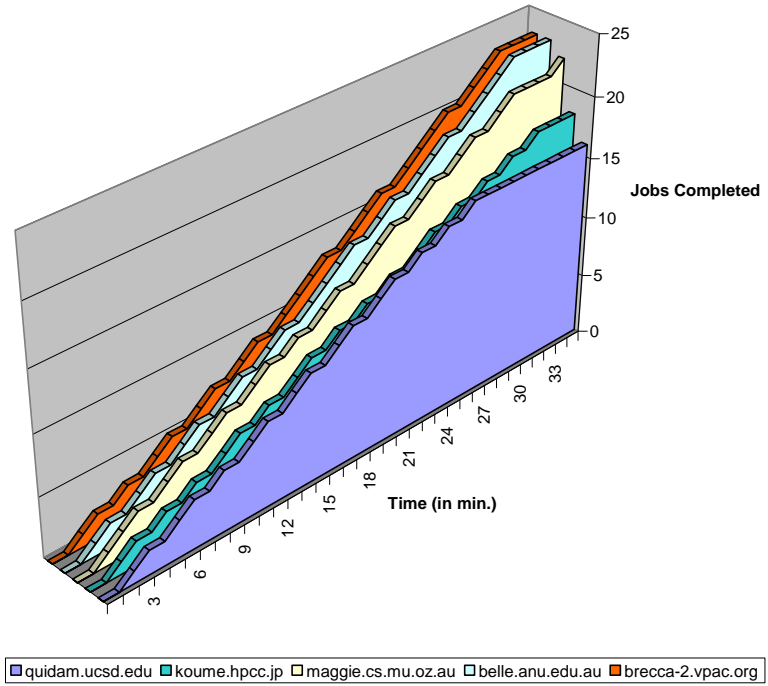

**Figure 12. A plot of the number of jobs completed on different resources versus the time.**

## 6   Related Work

In addition to its implementation based on service-oriented architecture using state-of-the-art technologies, Alchemi has a number of distinguished features when compared to related systems. Table 2 shows a comparison between Alchemi and some related systems such as Condor, SETI@home, Entropia, GridMP, and XtermWeb.

Condor [19] is the product of Condor Project at the University of Wisconsin-Madison. Condor can be used to manage a cluster of dedicated compute nodes (such as a "Beowulf" cluster). In addition, unique mechanisms enable Condor to effectively harness wasted CPU power from otherwise idle desktop workstations. Condor provides a job queuing mechanism, scheduling policy, priority scheme, resource monitoring, and resource management. Users submit their serial or parallel jobs to Condor, Condor places them into a queue, chooses when and where to run the jobs based upon a policy, carefully monitors their progress, and ultimately informs the user upon completion. It can handle both Windows and UNIX class resources in its resource pool.

The Search for Extraterrestrial Intelligence (SETI@home) [9][14] project borrows your computer when you are not using it to achieve larger, non-personal scientific goals. It aims towards analysis of scientific data on desktop clients. It is mainly based on the notion of peer to peer computing with centralized control.

Entropia [17] facilitates a Windows desktop grid system by aggregating the raw desktop resources into a single logical resource. Its core architecture is centralized in which a central job manager administers various desktop clients. The node manager provides a centralized interface to manage all of the clients on the Entropia grid, which is accessible from anywhere on the enterprise network.

XtermWeb [16] is P2P [15][18] project developed at University of Paris-Sud, France. It implements three distinct entities, the coordinator, the workers and the clients to create a so-called XtermWeb network. Clients are software instances available for any user allowed to submit tasks to the XtermWeb network, it submits tasks to the coordinator, providing binaries and optional parameter files and permits the end user to retrieve his results. Finally, the workers are software part spread among volunteer hosts to compute tasks.

|  | Alchemi | Condor | SETI@home | Entropia | XtermWeb | Grid MP |
|---|---|---|---|---|---|---|
| Architecture | Hierarchical | Hierarchical | Centralized | Centralized | Centralized | Centralized |
| Web Services Interface for Cross-Platform Integration | Yes | No | No | No | No | No |
| Implementation Technologies | C#, Web Services on Windows + .NET Framework | C | C++, Win32 | C++, Win32 | Java, Linux | C++, Win32 |
| Multi-Clustering | Yes | Yes | No | No | No | No |
| Global Grid Brokering Mechanism | Yes (via Gridbus Broker) | Yes (via Condor-G) | No | No | No | No |
| Thread Programming Model | Yes | No | No | No | No | No |

Table 2. Comparison of Alchemi and some related desktop grid systems.

The United Devices Grid MP (MP) [22] platform enables organizations to dramatically enable organizations to enhance the performance of their compute intensive applications by aggregating compute resources already available on their corporate network. Grid MP basically has a centralized architecture, where a Grid MP service acting as a manager accepts jobs from the user, schedules them on the resources having pre-deployed Grid MP agents. The Grid MP agents can be deployed on clusters, workstations or desktop computers. Grid MP agents receive jobs and execute them on resources, advertise their resource capabilities on Grid MP services and return results to the Grid MP services for subsequent collection by the user.

## 7 Summary and Future Work

We have discussed a .NET-based grid computing framework that provides the runtime machinery and object-oriented programming environment to easily construct desktop grids and develop grid applications. Its integration into the global cross-platform grid has been made possible via support for execution of grid jobs via a web services interface and the use of a broker component.

We plan to extend Alchemi in a number of areas. Firstly, support for additional functionality via the API including inter-thread communication is planned. Secondly, we are working on support for multi-clustering with peer-to-peer communication between Managers. Thirdly, we plan to support utility-based resource allocation policies driven by economic, quality of services, and service-level agreements. Fourthly, we are investigating strategies for adherence to OGSI standards by extending the current Alchemi job management interface. This is likely to be achieved by its integration with .NET-based low-level grid middleware implementations (e.g., University of Virginia's OGSI.NET [32]) that conform to grid standards such as OGSI (Open Grid Services Infrastructure) [24][31]. Finally, we plan to provide data grid capabilities to enable resource providers to share their data resources in addition to computational resources.